\documentclass[aps,prapplied,superscriptaddress,groupedaddress,twocolumn]{revtex4-2}

\usepackage{amsthm,amsmath}
\usepackage[utf8]{inputenc} 

\usepackage{graphicx}
\usepackage{dcolumn}
\usepackage{bm}
\usepackage{mathtools}
\usepackage{gensymb}
\usepackage[colorlinks=true]{hyperref}
\usepackage[dvipsnames]{xcolor}
\usepackage[mathlines]{lineno}

\newcommand\myshade{85}
\colorlet{mylinkcolor}{Red}
\colorlet{mycitecolor}{Blue}
\colorlet{myurlcolor}{Gray}

\hypersetup{
  linkcolor  = mylinkcolor!\myshade!black,
  citecolor  = mycitecolor!\myshade!black,
  urlcolor   = myurlcolor!\myshade!black,
  colorlinks = true,
}

\begin{document}

\title{Readout of quantum devices with a sideband microwave interferometer immune to systematic noise}

\author{N. Crescini}
	\altaffiliation[Present address: ]{Univ.Grenoble Alpes, CNRS, Grenoble INP, Institut N\'eel, 38000 Grenoble, France}
	\affiliation{IBM Research Europe, S\"aumerstrasse 4, 8803 R\"uschlikon, Switzerland}
	
\author{E. G. Kelly}
	\affiliation{IBM Research Europe, S\"aumerstrasse 4, 8803 R\"uschlikon, Switzerland}

\author{G. Salis}
	\email{gsa@zurich.ibm.com}
	\affiliation{IBM Research Europe, S\"aumerstrasse 4, 8803 R\"uschlikon, Switzerland}
	
\author{A. Fuhrer}
	\affiliation{IBM Research Europe, S\"aumerstrasse 4, 8803 R\"uschlikon, Switzerland}

\begin{abstract} 
The accuracy of microwave measurements is not only critical for applications in telecommunication and radar, but also for future quantum computers. Qubit technologies such as superconducting qubits or spin qubits require detecting minuscule signals, typically achieved by reflecting a microwave tone off a resonator that is coupled to the qubit. Noise from cabling and amplification, e.g. from temperature variations, can be detrimental to readout fidelity. 
We present an approach to detect phase and amplitude changes of a device under test based on the differential measurement of microwave tones generated by two first-order sidebands of a carrier signal. The two microwave tones are sent through the same cable to the measured device that exhibits a narrow-band response for one sideband and leaves the other unaffected. The reflected sidebands are interfered by down-conversion with the carrier. By choosing amplitude and phases of the sidebands, suppression of either common amplitude or common phase noise can be achieved, allowing for fast, stable measurements of frequency shifts and quality factors of resonators.
Test measurements were performed on NbN superconducting resonators at 25\,mK to calibrate and characterise the experimental setup, and to study time-dependent fluctuations of their resonance frequency.
\end{abstract}

\maketitle

\section{Introduction}
\label{sec:1}
\begin{figure*}[t!]
    \centering
    \includegraphics[width=.85\textwidth]{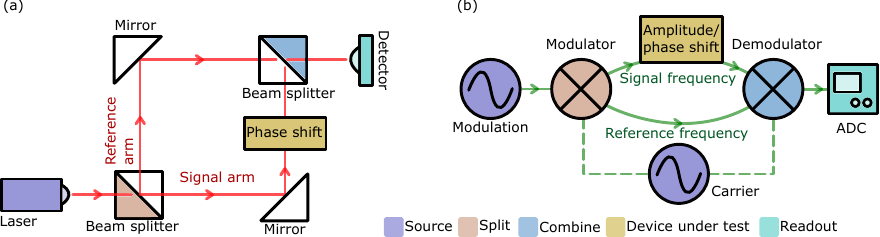}
    \caption{Comparison between (a) an optical Mach-Zehnder interferometer and (b) a microwave sideband interferometer as the one presented in this work. In (a), a laser is split into two rays that follow different paths and eventually interfere on a detector. In (b) the signal follows the same path but contains two sidebands of different frequency, where only one is affected by the device under test. Down-conversion with the carrier signal yields the measured interference signal.}
    \label{fig:1}
\end{figure*}

Interferometric measurements are among the most sensitive and well studied in experimental physics. Their application ranges from fundamental physics and metrology to quantum information~\cite{PhysRevLett.80.3181,Hudelist2014,10.1117/12.621581,Martin2020}, and they are of transversal interest for many other fields of research~\cite{steel_interferometry_1983,hariharan_optical_2003}. In an optical interferometer, the probe and reference laser beams follow distinct paths and their interference is indicative of relative variations between the two paths, see Fig.\,\ref{fig:1}(a). Microwave interferometers can also be realized in this way~\cite{pozar_microwave_2011} and were used to measure frequency variations of devices such as resonators~\cite{rubiola2008phase,738292,721162,Regal2008,doi:10.1063/1.1347971,doi:10.1063/1.4797470,doi:10.1063/1.3648134}. For qubit readout in solid state quantum processors~\cite{Crippa2019,PhysRevApplied.13.024019,PRXQuantum.2.010353,Martin2020,doi:10.1063/1.3273372,PhysRevB.84.064517,PhysRevLett.121.090502,PhysRevLett.123.190502,Lisenfeld2016} or to measure frequency variations and noise in resonators used for kinetic inductance detectors~\cite{doi:10.1146/annurev-conmatphys-020911-125022, Day2003,Baselmans2008}, standard single side band (SSB) detection is employed to convert a variation of the microwave signal to baseband. In that scheme, a modulation signal is up- and down-converted in frequency allowing for sensitive measurements at high frequency with relatively inexpensive baseband instrumentation.

One of the advantages of interferometric measurements is their insensitivity to common mode noise or long-term drifts common to both interfering paths. Measurements where the two paths are likely to experience differential noise -- e.g. when a microwave signal senses a sample in a cryostat and is then referenced to a carrier signal routed outside the cryostat -- cannot benefit from this suppression, and the signal will be affected by variations in temperature, losses of microwave components or amplification noise specific to one path.

Here, we demonstrate how the two paths of an interferometer in the microwave domain can be substituted by two different frequencies, thus using a bichromatic signal along the same physical path (e.\,g. the same cable going into and out of a cryostat). In this case a relative phase or amplitude shift between the two microwave tones produces the desired signal, whereas noise in amplitude or phase common to the two tones can be suppressed. 

The basic idea to separate the two interfering signals in the frequency domain is shown in Fig.\,\ref{fig:1}(b). Our approach is based on sideband modulation of a carrier signal, and is thus referred to as a sideband microwave interferometer (SMI).
We describe the implementation and characterisation of the SMI, provide a theoretical model and experimentally verify the suppression of common-mode phase noise and amplitude noise at two different operation points of the interferometer. The setup is then used to probe the response of a coplanar waveguide resonator at millikelvin temperatures,
allowing rapid microwave measurements of fluctuations of the resonator's frequency and quality factor induced by two-level-systems (TLSs). 
The experiment is implemented with a signal generator, a splitter, two mixers, and a lock-in amplifier. 
The bandwidth is limited by the shortest lock-in integration time, and in the present case can extend up to tens of MHz, allowing for the measurement of fast variations although at the expense of the detection sensitivity.

In the following Section we provide a detailed explanation of the measurement scheme and its experimental implementation. Section \ref{sec:3} shows the operation of the SMI, and characterization of its noise suppression properties. In Section \ref{sec:4} we conclude and provide perspectives on improvements with which we plan to advance our scheme.

\section{Measurement scheme}
\label{sec:2}
The general idea [see Fig.\,\ref{fig:1}(b)] is to modulate a microwave carrier signal to generate two sidebands. These are reflected off or transmitted through a device under test, and then demodulated to make them interfere. In contrast to conventional SSB detection~\cite{doi:10.1063/1.4981390}, the interference allows for the detection of relative phase or amplitude changes between the two sidebands, which can be exploited to get a first-order suppression of the common-mode noise introduced e.\,g. in the cable by thermal expansion or vibrations. For instance, one sideband can be tuned to the frequency of a resonator and be used as a probe, while the other one may lie several linewidths away, acting as a reference. 
The parameters of the SMI can be tuned to obtain an output signal that is proportional to a constructive or destructive interference of the two reflected or transmitted sidebands, thus reducing either common-mode phase or amplitude noise.

The definitions of the various frequencies used in this work are reported in Table \ref{tab:1} in Appendix\,\ref{app:details}, and are hereafter used to detail the theoretical description of this method.
    
\begin{figure}[b!]
\includegraphics[width=.4\textwidth]{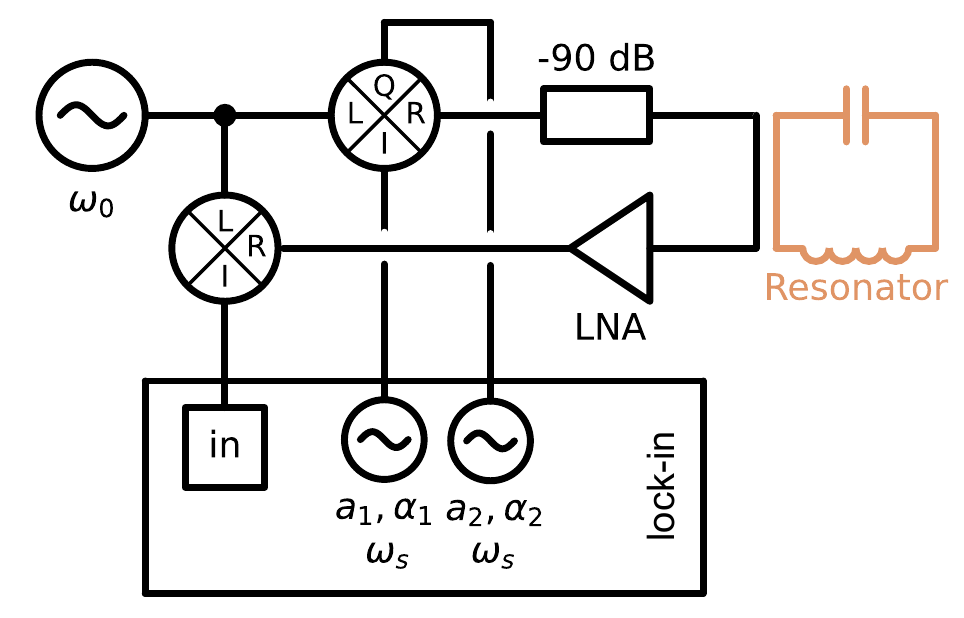}
\caption{Schematics of the experimental setup. Here, $-90\,\mathrm{dB}$ indicates the estimated attenuation of the microwave lines in the cryostat, LNA denotes the low noise amplification chain both at cryogenic and at room temperature.}
\label{fig:2}
\end{figure}

Fig.\,\ref{fig:2} shows the setup used in this work. The carrier signal proportional to $e^{i\omega_0 t}$ is modulated in an IQ mixer with $I$ and $Q$ signals $a_1e^{i\omega_s t + i\alpha_1}$ and $a_2e^{i\omega_s t + i\alpha_2}$ generated by a lock-in amplifier. This creates two first-order sidebands at frequencies $\omega_{1,2}=\omega_0\pm \omega_s$. The amplitudes and phases of the two sidebands can be selected by setting the modulation parameters $a_1$, $a_2$, $\alpha_1$ and $\alpha_2$. 
More specifically, the sidebands are proportional to $a_1 e^{i\omega_1t + i\alpha_1}-a_2 e^{i\omega_1t + i\alpha_2}$ and $a_1 e^{i\omega_2t - i\alpha_1}+a_2e^{i\omega_2t - i\alpha_2}$. 

In our case the signal is then attenuated in multiple steps (by a total of 90\,dB) and sent to a coplanar waveguide resonator in a dilution refridgerator. The reflected signal is amplified by a low-noise cryogenic amplifier (LNA) at 4\,K. A second amplification stage at room temperature further increases the signal strength resulting in a combined amplification of 70\,dB. The output signal is then down-converted by mixing it with the carrier signal $e^{i \omega_0 t + i\delta}$ at a second mixer (accounting for a phase difference $\delta$ of the carrier signal between the two mixers). The output of the latter is finally measured using the lock-in amplifier yielding in-phase and quadrature components $x$ and $y$.

For simplicity we assume that the device under test only affects the signal at frequency $\omega_1$, in this sense the two sidebands can be referred to as probe ($\omega_1$) and reference ($\omega_2$). We assume that the device under test shifts the phase of the probe sideband by $\phi$ and reduces its amplitude by a factor $\xi$. Neglecting the terms that oscillate with $2\omega_0$, the output signal of the SMI that is measured by the lock-in amplifier is
\begin{align}
    \begin{split}
    s_{\mathrm{out}}(t) & = a_1 e^{i\omega_s t+i\alpha_1+i\delta} + \xi a_1 e^{i\omega_s t + i\alpha_1 + i\phi - i\delta}  \\
                       & + a_2 e^{i\omega_s t + i\alpha_2 + i\delta} - \xi a_2 e^{i\omega_s t + i\alpha_2 + i\phi - i\delta}.
\label{eq:2}
    \end{split}
\end{align}

\begin{figure}[t!]
\centering
\includegraphics[width=.48\textwidth]{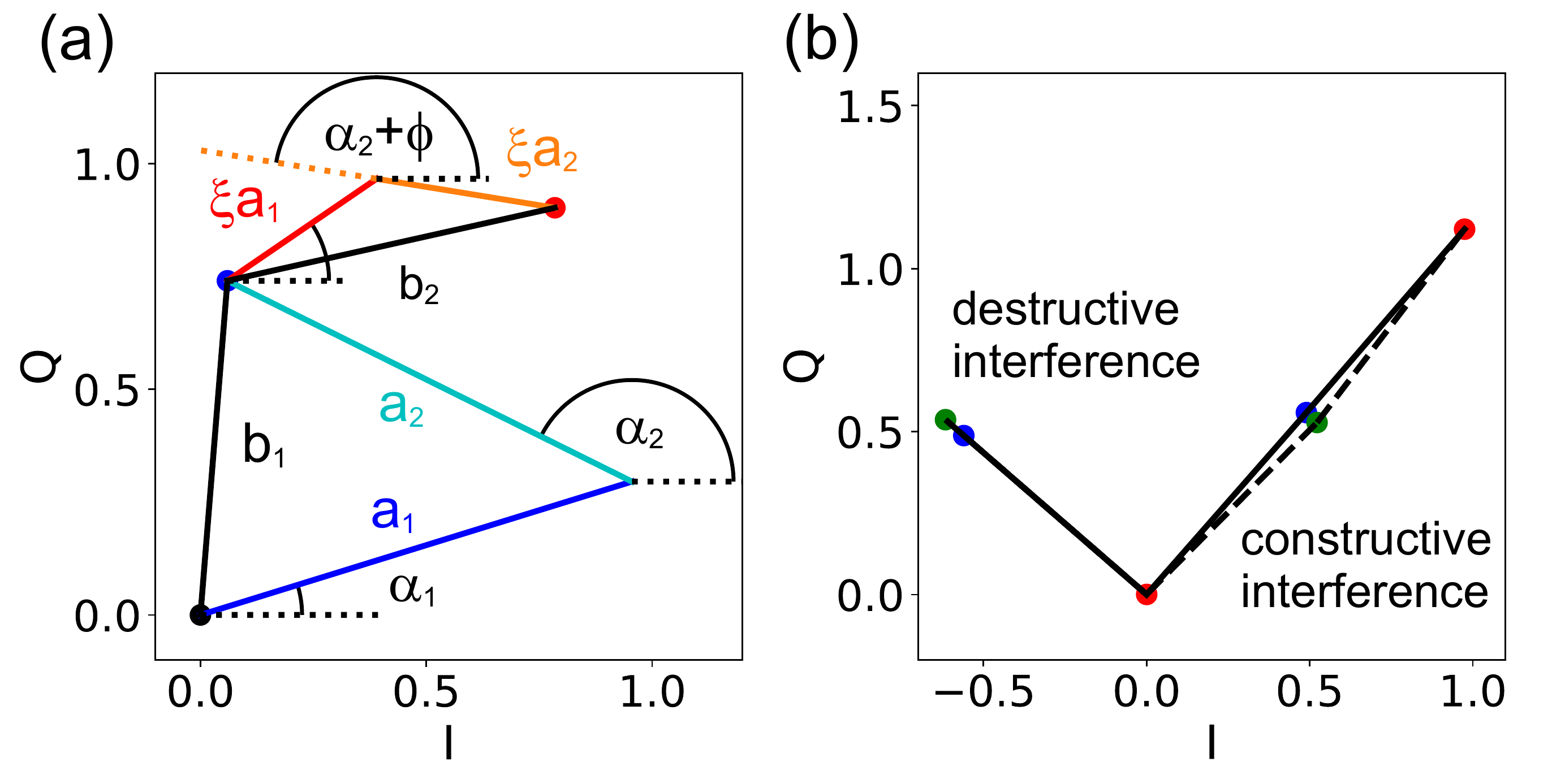}
\caption{Interference of the down-converted signal in the complex plane. (a) Four terms in Eq.~\ref{eq:2} of the up- and down-converted signal after acquiring a phase shift $\phi$ and an attenuation $\xi$ at the lower sideband frequency, where the red dot indicates the resulting interference of the two terms $b_1$ and $b_2$. Note that we assume $\delta = 0.0$. (b) Two situations with parameters adjusted to obtain constructive interference, where the common phase noise is suppressed, or destructive interference, where common amplitude noise is suppressed. Blue dots correspond to positions of $b_1$ without common-mode offsets, green dots to those with common-mode phase increase of 0.06 rad (for constructive interference, $\delta = -0.64$) and common-mode amplitude increase of 10 percent (for destructive interference $\delta=0.94$). Used parameters values: $a_1=a_2=1.0$, $\alpha_1=0.3$, $\alpha_2=2.68$, $\xi=0.4$ and $\phi=0.3$.}
\label{fig:3}
\end{figure}

The in-phase and quadrature components of the terms oscillating at $\omega_s$ in Eq.\,(\ref{eq:2}) are visualized in Fig.~\ref{fig:3}(a). The total signal can be understood as two interfering terms, $b_1 e^{i\beta_1}=a_1 e^{i\alpha_1+i\delta}+a_2 e^{i\alpha_2 + i\delta}$ and  $b_2 e^{i\beta_2}=\xi e^{i \phi}(a_1 e^{i\alpha_1 - i\delta}-a_2 e^{i\alpha_2 - i\delta})$. The amplitudes $b_1$ and $b_2$ are given by

\begin{equation}
b_1^2=a_1^2+a_2^2+2a_1a_2\cos\left(\alpha_1-\alpha_2\right)
\label{eq:2a}
\end{equation}
and
\begin{equation}
b_2^2=\xi^2\left(a_1^2+a_2^2-2a_1a_2\cos\left(\alpha_1-\alpha_2\right)\right).
\label{eq:2b}
\end{equation}

By tuning the parameters of the modulation signals ($a_1$, $\alpha_1$, $a_2$, $\alpha_2$ and $\delta$) we can achieve identical amplitudes for the two interfering terms, $b_1 = b_2$, and an arbitrary phase $\beta_1 - \beta_2$ between them. The parameters for same amplitudes are found by equating Eq.~\ref{eq:2a} with Eq.~\ref{eq:2b}, yielding the condition $\cos(\alpha_1-\alpha_2)= (\xi^2-1)(a_1^2+a_2^2)/((1+\xi^2)(2a_1a_2))$. The phase $\beta_1 - \beta_2$ can be tuned by adapting $a_1$, $a_2$ and either $\alpha_1 - \alpha_2$ or $\delta$.

There are two phases of specific interest, geometrically illustrated in Fig.~\ref{fig:3}(b): for $\beta_1 - \beta_2=0$, we achieve constructive interference, and common mode phase noise at the two sideband frequencies is suppressed. For $\beta_1 - \beta_2=\pi$, we achieve destructive interference, where common noise in the amplitude of the two sidebands is canceled. For both, destructive and constructive interference, the signal demodulated by the lock-in amplifier senses $\xi$ and $\phi$, in a similar way as in conventional SSB detection schemes, but here with the added advantage of suppressed common-mode noise.

\begin{figure}[b!]
\centering
\includegraphics[width=0.5\textwidth]{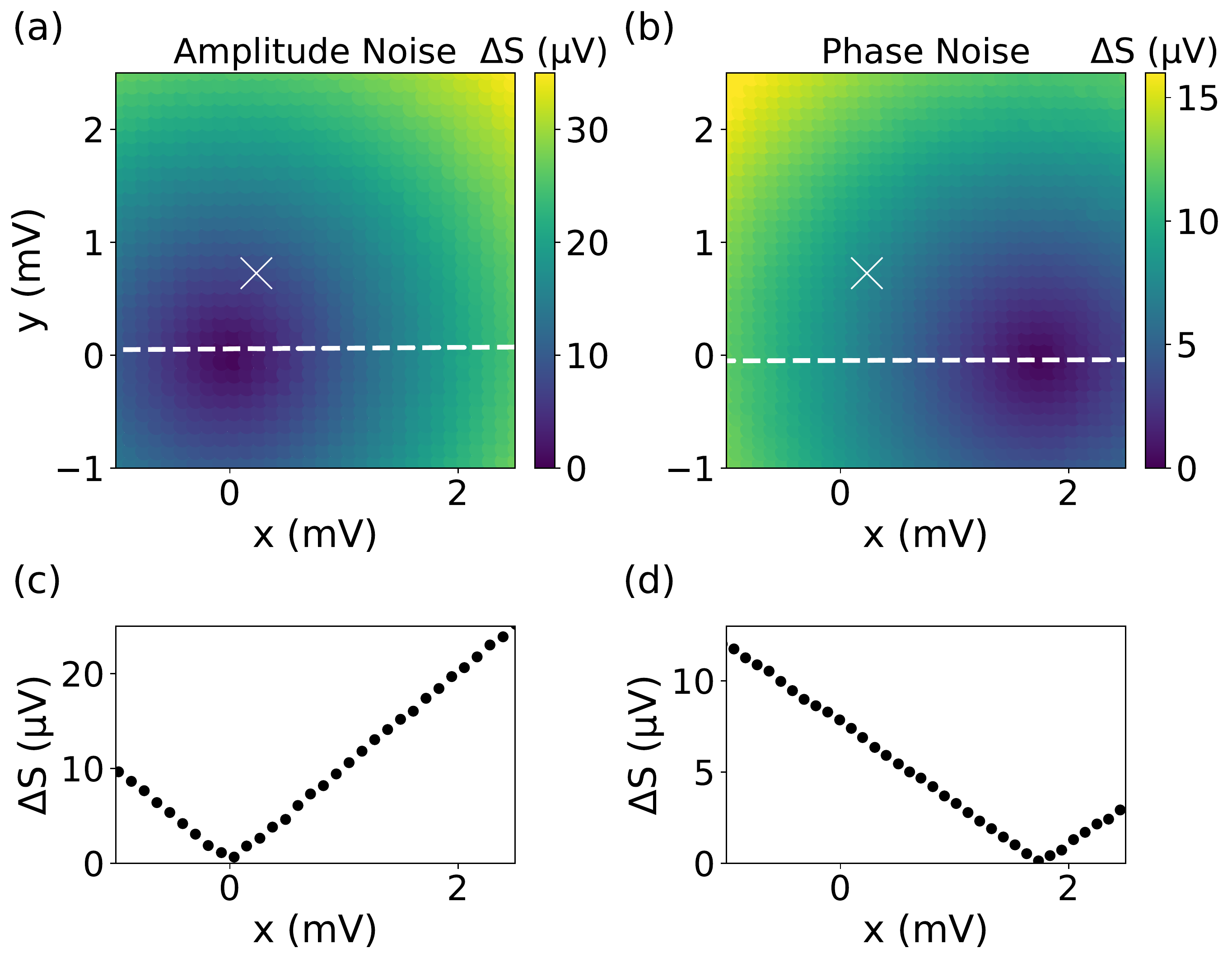}
\caption{Measured signal variation $\Delta S$ for a common modulation of (a) the  amplitudes of the two sidebands by one percent and (b) the phase of the two sidebands by 0.01 radians. These quantities are shown at different working points in the IQ plane of the SMI obtained by varying $a_2$ and $\alpha_2$ for fixed $a_1$ and $\alpha_1$. The white cross indicates the single sideband point in the IQ plane. At the optimal working points, the influence of either noise source linearly approaches zero, as seen in the cross sections in (c) and (d) taken at the dashed white lines in (a) and (b).}
\label{fig:4}
\end{figure}

\section{Test and calibration of the sideband interferometer}
\label{sec:3}

\subsection{Phase and amplitude noise rejection}
\label{sec:noiserej}
To verify the predicted common-mode amplitude-noise and phase-noise rejection, we use the interferometric scheme reported in Fig.\,\ref{fig:2} but replace the resonator with a variable attenuator and a variable phase shifter. We determine the optimal rejection points for both noise sources by applying either a controlled attenuation or a controlled phase shift to the up-converted signal. In Fig.~\ref{fig:4}, we compare the magnitude of noise sensitivity $\Delta S = \sqrt{\Delta x^2 + \Delta y^2}$ as a function of varying $a_2$ and $\alpha_2$ (at fixed $a_1$ of 10\,mV  and $\alpha_1$ of 330 degrees). $\Delta x$ and $\Delta y$ correspond to the changes of the I and Q components $x$ and $y$ of the demodulated signal when varying the common-mode amplitude by one percent or the common-mode phase by 0.01 rad. The detailed experimental setup is presented in Appendix\, \ref{app:details}. 

Common amplitude noise is suppressed at $x=y=0.0$\,mV [see Fig.~\ref{fig:4}(a)] as predicted in Sec.\,\ref{sec:2} and because of destructive interference of the two sidebands. We reach this point at $a_2$ = 43.7\,mV   and $\alpha_2=$246.4\degree. The linecut of $\Delta S$ versus $x$ at $y=0.0$\,mV in Fig.~\ref{fig:4}(c) shows that $\Delta S$ linearly decreases towards the rejection point. Similiarly, the common-mode phase noise [see Fig.~\ref{fig:4}(b)] is linearly suppressed towards its rejection point at $x=1.7$\,mV and $y=0.0$\,mV, see Fig.~\ref{fig:4}(d). This point corresponds to  $a_2$ = 23.2\,mV and $\alpha_2$ = 57.2\degree. The values in Figs.~\ref{fig:4}(c) and (d) can be compared to the noise sensitivity at the SSB configuration [$a_2=11$\,mV and $\alpha_2=$61\degree, crosses in Fig.~\ref{fig:4}(a) and (b)], where $\Delta S$ is $7.6\,\mathrm{\mu V}$ for both one percent of amplitude noise or 0.01 rad of common-mode phase change. These results quantitatively demonstrate that the SMI can to first order suppress either form of common-mode noise by adjusting amplitude and phase of the two lock-in outputs. We expect that the ultimate limit of such suppression is given by the stability of phases and amplitudes of the lock-in outputs and by that of the carrier signals used for the up- and down-conversion.

\subsection{Resonator measurements}
\label{sec:smi_res}

To test the interferometer, we use a high kinetic inductance coplanar waveguide resonator in a hanger geometry, fabricated from a thin film of sputtered NbN with a nominal thickness of 10\,nm on an intrinsic silicon substrate.
The resonator has a length of $2\,\mathrm{mm}$. Its resonance frequency is $f_r=6.16\,\mathrm{GHz}$, the loaded and coupling quality factors are $Q_l \simeq 8.0 \times 10^3$ and $Q_c\simeq9.8 \times 10^3$, and the resonator has an impedance $Z_r \simeq 316\,\mathrm{\Omega}$. Since the low-power internal quality factor is of order $Q_\mathrm{i}\simeq5\times10^{4}$, the resonator is overcoupled. Measurements are performed at 25\,mK, the base temperature of a dilution refrigerator. The reader is referred to the appendices for more details, which show the measured probe signal (microwave output) of the SMI and the phase and amplitude response of the resonator under study.

If the frequency of the probe sideband is set to the resonator's center frequency $\omega_r/2\pi$, a small variation of $\omega_r$ by $\delta_r\ll\gamma_r$ mostly affects the phase $\phi$ of the reflected sideband. Here, $\gamma_r$ is the resonator linewidth. With the SMI probing at a fixed frequency, only frequency shifts $\delta_r < \gamma_r$ can be resolved. For larger $\delta_r\gg\gamma_r$, the signal exceeds the measurement bandwidth (MBW) and changes in $\phi$ or $\xi$ become very small.

\begin{figure}
\centering
\includegraphics[width=.4\textwidth]{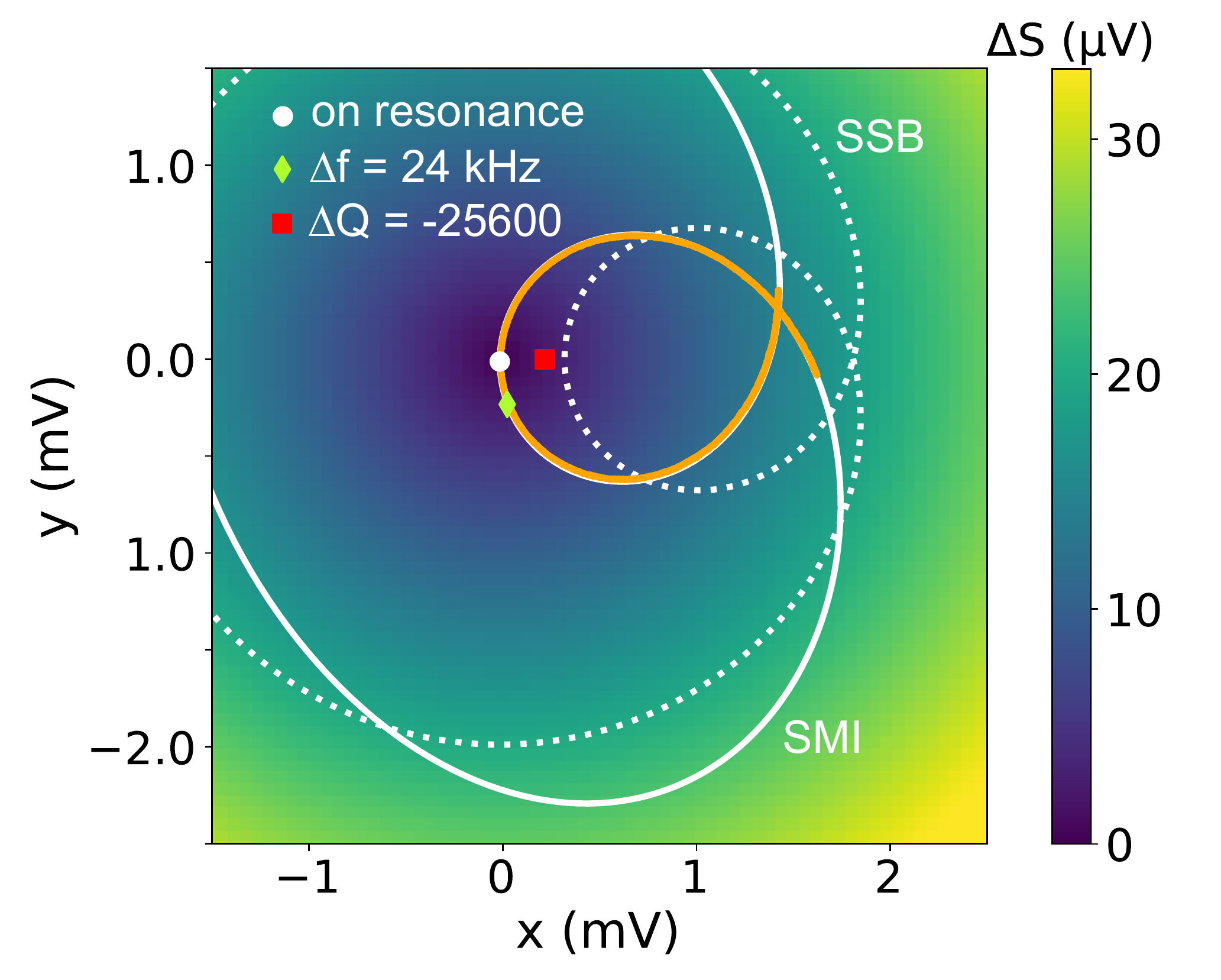}
\caption{Spectroscopy of a resonator and comparison of SSB and SMI signal in the IQ plane. The dashed curve shows the calculated reflection from the resonator in the IQ plane for normal single-sideband detection, and makes the link to the VNA data shown in  Fig.\,\ref{fig:7} in Appendix\,\ref{app:details}. The orange line is the measured data in amplitude noise rejection configuration of the SMI. The solid line is the fitted theoretical curve of the interferometric readout. For noise measurements, the probe frequency was set close to the resonance frequency of the resonator (white dot). Small variations in resonator frequency (green diamond, frequency shift of 24 kHz) or resonator quality factor (red dot, reduction of internal quality factor from $Q_\mathrm{i}$ to $Q_\mathrm{i}/2$) correspond to orthogonal directions in the IQ plane. The colorscale indicates the magnitude $\Delta S$ of changes of the signal in the IQ plane when assuming 1 percent of common-mode amplitude noise on both sidebands.}
\label{fig:5}
\end{figure}
    
 To determine absolute frequency shifts of the resonator, it is necessary to relate the SMI signal to the resonance lineshape. The latter is obtained from SSB spectroscopy by sweeping the frequency of the carrier $\omega_0$ in an interval $\simeq \gamma_r$, with $\omega_1$ close to the resonator frequency, and independently with a vector network analyzer (VNA). In Fig.~\ref{fig:5} this is shown together with the signal where the SMI is tuned to the amplitude-noise suppression point at $\omega_1=\omega_r$ (orange curve). Note that the electrical phase delay (i.e. a background phase component linearly varying with frequency) leads to a circular (elliptical) background signal in the SSB (SMI) configuration, respectively. 
If probed on resonance, small shifts of the resonator frequency affect mostly the phase $\phi$ of the reflected probe sideband, whereas variations in the quality factor change the reflected amplitude $\xi$, changing the SMI signal in the orthogonal direction in the IQ $(x, y)$ plane of the demodulated signal. In Fig.~\ref{fig:5} we indicate the calculated SMI signal for a resonator with reduced internal quality factor (red square) and with a shifted resonance frequency (green diamond).

\subsection{Study of resonator noise}
\label{sec:tls}
Noise in microwave resonators usually originates from the presence of TLSs in amorphous materials or interfaces  of these devices~\cite{PhysRevLett.109.157005,Muller2019,PhysRevB.92.035442,Faoro2015InteractingTM}. Due to the relevance for improving coherence of quantum devices, this is the object of an intense research activity~\cite{doi:10.1063/1.2711770,doi:10.1063/1.2937837,PhysRevB.80.132501,5067054,doi:10.1063/1.3309754,Lisenfeld2015ObservationOD,Burnett2015AnalysisOH,doi:10.1063/1.4919761,Neill2013FluctuationsFE,PhysRevLett.119.264801}.
Techniques that track frequency fluctuations~\cite{doi:10.1119/1.1286663} can shed light on the properties of TLSs~\cite{doi:10.1063/1.3648134,6785973,niepce2020stability}, and have brought on new possibilities for studying their distribution, origin and stability~\cite{Burnett2014EvidenceFI,Burnett2013SlowNP,Skacel2015ProbingTD,doi:10.1063/1.5001920,doi:10.1063/1.5053660,Burnett2018,PhysRevApplied.11.044014,niepce2020stability}.
In this context the SMI, that allows suppression of common mode noise, can be a powerful tool to gain further insights into the underlying TLS physics.

\begin{figure*}
\centering
\includegraphics[width=.75\textwidth]{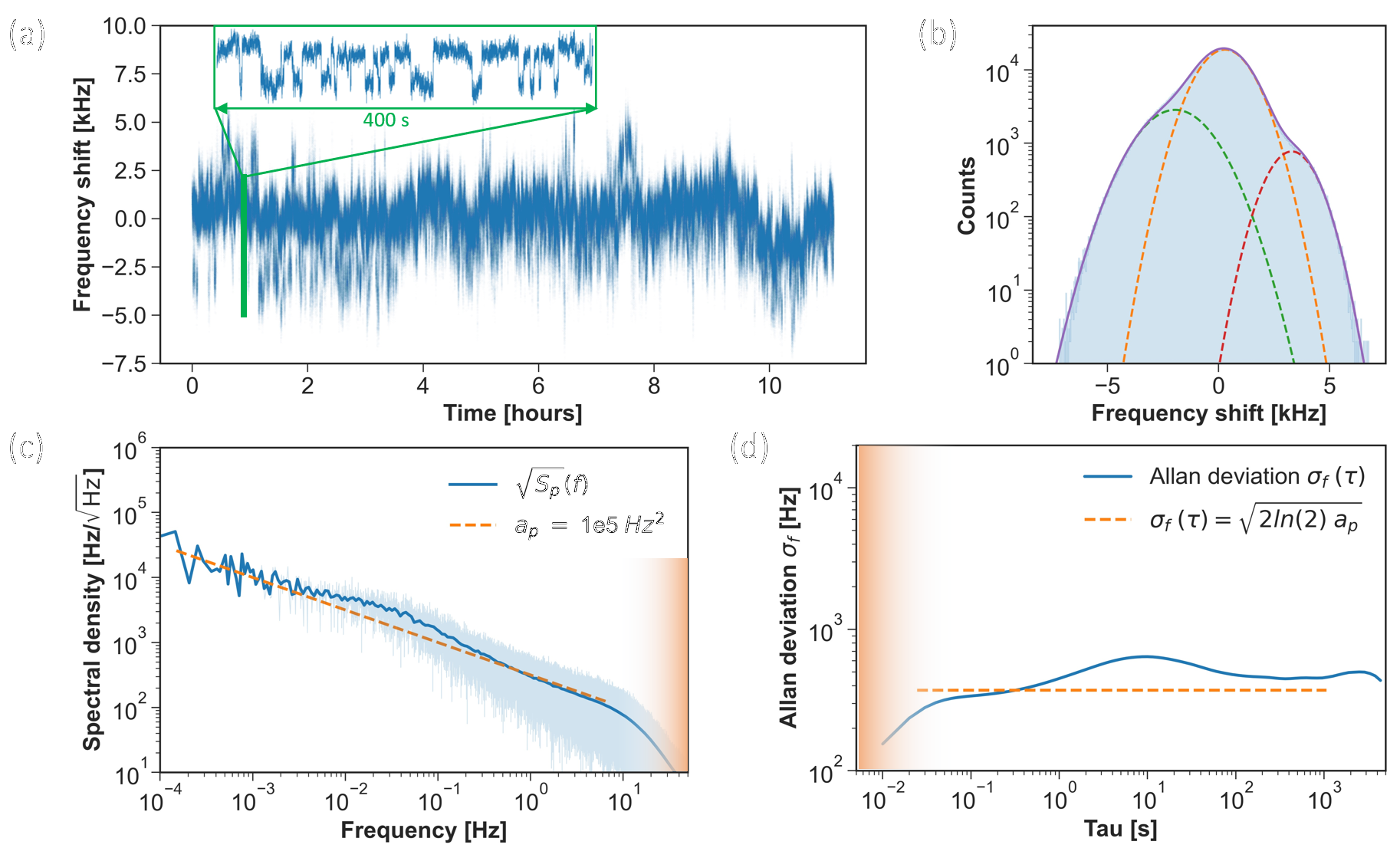}
\caption{SMI measurement of frequency fluctuations of a test resonator. (a) time trace, (b) histogram, (c) spectral density $\sqrt{S_p}$ and (d) Allan deviation $\sigma_f(\tau)$ of the same data set. The average photon number in this measurement is $\langle n \rangle \simeq 8$. The dashed brown lines indicate the expected ${a_p}/{f}$ behavior of the power spectrum $S_p$ and the brown shading indicates the limit imposed by the measurement bandwidth. }
\label{fig:6}
\end{figure*}

With the SMI tuned to the amplitude noise rejection point, we have measured resonator frequency noise fluctuations for more than 11 hours at a time and an example trace is reported in Fig.\,\ref{fig:6}(a). Frequency fluctuations of the resonator were obtained from the SMI signal using a calibration as discussed in Section\,\ref{sec:smi_res} and using a linear interpolation between SMI phase and resonator frequency. The detected noise increases by more than a factor of $100$ with respect to the signal measured off resonance, and is thus dominated by the resonator's fluctuations. The average photon occupation number was calculated using~\cite{doi:10.1063/1.4919761}
\begin{equation}
    \langle n \rangle = \frac{E_r}{\hbar \omega_r} = \frac{2}{\hbar \omega_r^2} \frac{Q_l^2}{Q_c} \frac{50\,\Omega}{Z_r}P_\mathrm{in},
    \label{eq:3}
\end{equation}
where $P_\mathrm{in}$ is the input power of the on-resonance sideband. In this measurement the resonator is studied with $\langle n \rangle\simeq 8$.
The sampling frequency is set to 100\,Hz and the lock-in integration time is 10\,ms.

The time trace of the resonator frequency shown in Fig.\,\ref{fig:6}(a) displays clear random telegraph signals (see inset) typically associated with TLSs noise.  
The histogram (b), spectral density (c) and Allen deviation (d) put a slightly different spotlight on the TLSs properties. The histogram (b) of the time trace in (a) can be fit with three shifted Gaussians. This seems to indicate that a few slow TLSs dominate the disturbance of the resonator and lead to three large characteristic frequency shifts. In an ensemble of TLSs, a few near resonant fluctuators can be more strongly coupled to the resonator~\cite{Muller2019} giving rise to such discrete offsets. 
The noise spectrum $\sqrt{S_p}$ in (c) is obtained through a Fourier transform of the time trace after subtracting offsets and by application of a moving average in the frequency domain. Here, we expect to see an ensemble of TLSs show up as the typical $S_p = a_p/f$ dependence (dashed line). There are two obvious deviations from this behavior in the data. Above 10\,Hz we see that the noise is suppressed due to the measurement bandwidth limited by the lock-in integration time (brown shading). At around 20\,mHz we find more noise than expected from the $1/f$ background. We argue that this is due to a characteristic timescale of the dominant TLSs that contribute to an additional broad peak in the spectrum \cite{niepce2020stability}. A comparison with the 400\,s inset in (a) shows that the switching frequency of the dominant TLS is compatible with this value. Another, complementary way of visualizing this, is the Allan deviation $\sigma_f(\tau)$ shown in Fig.\,\ref{fig:6}(d). The Allan deviation for a $a_p/f$ power spectrum should be constant at a level of $\sigma_f(\tau) = \sqrt{2\ln2\, a_p}$ (dashed line). We again find a peak centered around $\tau =1/(2\pi \,20\,{\mathrm{mHz}}) \approx 8$\,s and a reduction of the noise at small $\tau$. 
We also studied the dependence of these noise spectra and Allan deviations down to photon numbers $\langle n \rangle < 0.1$ where white noise dominates the spectrum for frequencies $>0.2$\,Hz and the low frequency fluctuation amplitude of the resonator increases to  $a_p = 4.0\cdot10^6$\,Hz$^2$ (see Appendix\,\ref{app:power}). These findings are in agreement with previously reported results, e.\,g.~\cite{doi:10.1063/1.3648134,niepce2020stability}.

This shows that the application of the SMI to the study of TLS noise is promising. The fact that variations in frequency and Q value of the resonator lead to orthogonal signals in the SMI readout (see Fig.~\ref{fig:5}) can be used to simultaneously track both frequency and quality factor shifts of a quantum device~\cite{6785973}. This allows the study of noise correlations in these two values, something that we will explore in a forthcoming work.

\section{Conclusions and perspectives}
\label{sec:4}
We describe and operate a microwave interferometer based on the modulation of a carrier resulting in two sidebands, which are used as probe and reference signals.
This setup measures relative phase or amplitude shifts between the probe and reference, therefore selectively eliminating common mode noise and enabling long-term stability in quantum device characterization.
We test our scheme by adding controlled common phase and amplitude variations which demonstrate the scheme's resilience to systematic noise. We use the SMI for the readout of a superconducting resonator at millikelvin temperatures, and measure its frequency noise induced by nearby TLS fluctuators. 

The ultimate sensitivity that can be reached with this setup is not thoroughly addressed in this work, but will be the subject of future analysis. A critically coupled resonator could be read out at lower power with the same signal-to-noise ratio. Exploiting this, one could test a resonator's noise with a much lower photon number $\langle n \rangle$ or increase the MBW to study higher frequency noise.
In the case of frequency shifts that exceed the linewidth of the resonator, a feedback loop with a PID controller could be implemented via software in the lock-in amplifier, in order to track the resonance frequency and extend the MBW of the present experiment. This addition does not necessarily reduce signal bandwidth, as the PID could be integrated digitally in the lock-in.

In the present work the potential of the SMI is demonstrated in a study of TLS noise affecting a coplanar-waveguide resonator, but we stress that its application is far more general. For instance, the properties of the SMI can be leveraged for the readout of qubits at lower power and reduced sensitivity to systematic sources of noise such as slow drifts induced by temperature fluctuations. Instead of using the two sidebands as probe and reference, the principle of sideband interference can be extended by using both sidebands as a probe e.g. of two different resonators each coupled to a qubit. This could enable parity readout of the two qubits with reduced common-mode noise.

\section*{Acknowledgements}
The authors would like to thank the Cleanroom Operations Team of the Binnig and Rohrer Nanotechnology Center (BRNC) for their help and support. We are also grateful to Stephan Paredes and Peter M\"uller for the experimental help, and to Clemens M\"uller, Giuseppe Ruoso and Leonardo Massai for fruitful discussions.
We gratefully acknowledge funding from Swiss National Science Foundation under grant number $200021\_188752$ and by the NCCR SPIN, funded by the SNSF under grant number 51NF40-180604.

\appendix
\section{Additional experimental details}
\label{app:details}
In the main text there are several frequency quantities and abbreviations which are important to tell apart. Table \ref{tab:1} sums them up and lists their name, symbol, a short description and their value. The first group of rows are frequencies and frequency shifts of the interferometer signals and of the resonator, while the second, separated by a line, describes the bandwidths of the readout system.

\begin{table*}
    \centering
    \begin{tabular}{ l c p{7cm} c }
    Name	&	Symbol	&	Description	& Indicative value	\\
    \hline
    \hline
    Carrier frequency		& $\omega_0/2\pi$ 		& Frequency of the microwave oscillator.		                & $6.16\,\mathrm{GHz}$ \\
    Modulation frequency	& $\omega_s/2\pi$ 		& Frequency of the phase and quadrature inputs of the IQ mixer.	& $10\,\mathrm{MHz}$ \\
    Sidebands frequencies	& $\omega_{1,2}/2\pi$	& Sideband microwave frequencies.			                    & $\omega_0\pm \omega_s$ \\
    Resonator frequency		& $\omega_r/2\pi$		& Average resonator frequency.				                    & $\omega_0/2\pi$ \\
    Resonator linewidth		& $\gamma_r/2\pi$	    & Average resonator linewidth.				                    & $\mathrm{100\,kHz}$ \\
    Resonator shift			& $\delta_r/2\pi$	    & Shift of the resonator frequency.			                    & $\sim\mathrm{10\,kHz}$ \\
    \hline
    Measurement band		& MBW		& Maximum frequency shift $\delta_r$ that the interferometer can detect.	& $<\gamma_r$ \\
    Sampling band		    & SBW		& Maximum frequency of $\delta_r$ which can be sampled, corresponding to the minimum lock-in $\tau$ of 30\,ns.	& $30\,\mathrm{MHz}$ \\
    Lock-in bandwidth	    & LBW		& Lock-in amplifier bandwidth, related to the analog-to-digital converter sampling rate of 1.8\,GSa/s.			& $600\,\mathrm{MHz}$ \\
    \hline
    \hline
    \end{tabular}
    \vspace{.1cm}
    \caption{Definition of the different frequencies which are relevant for this setup.}
    \label{tab:1}
\end{table*}

\begin{figure}
\includegraphics[width=.45\textwidth]{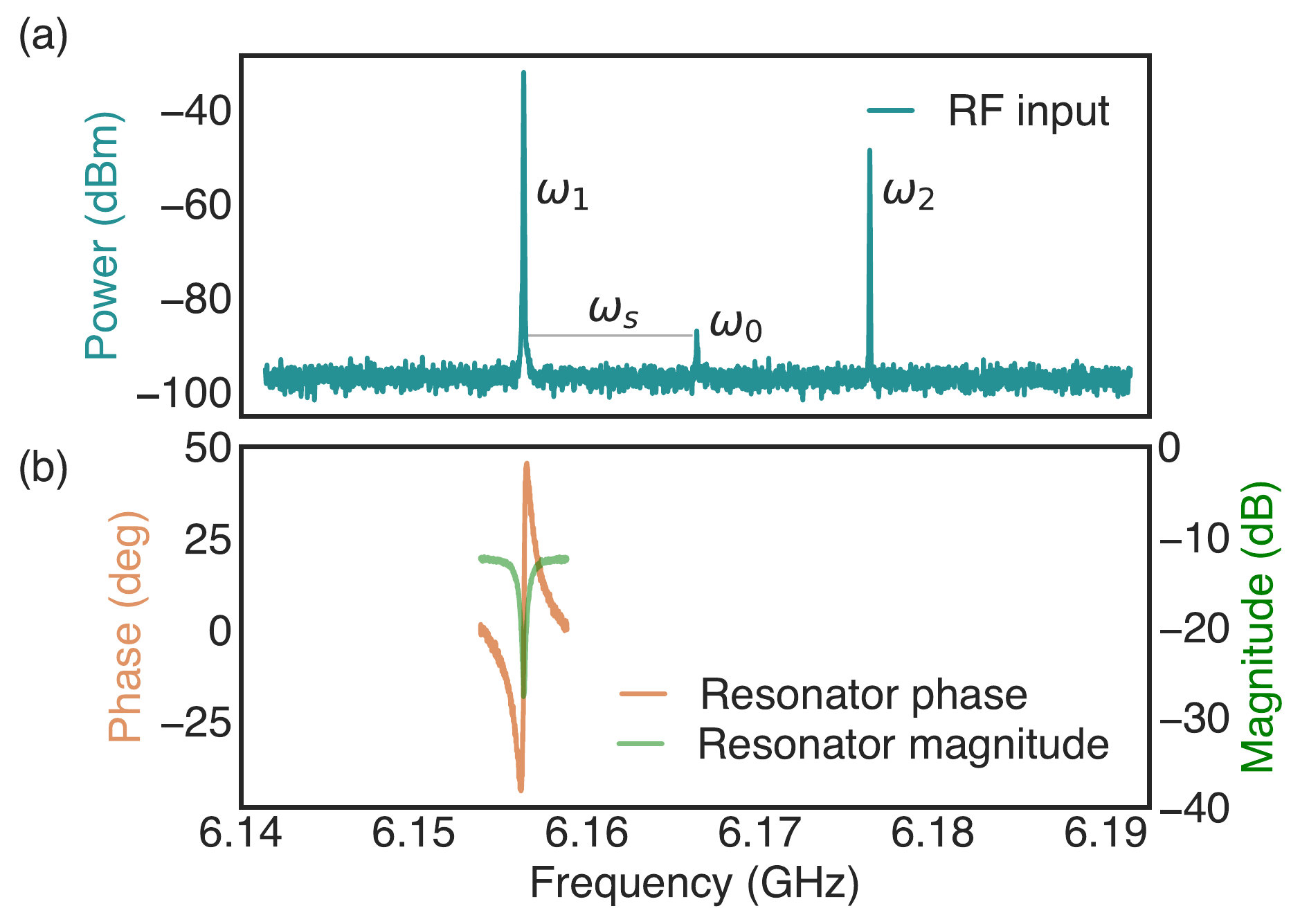}
\caption{Interferometer probe signal and resonator response. (a) Spectrum of the microwave IQ mixer output signal that is sent to the resonator in the dilution refridgerator, with probe sideband at $\omega_1$, reference sideband at $\omega_2$ and a residual carrier signal at $\omega_0$. (b) Phase and magnitude of the resonator lineshape that is probed by the sideband at $\omega_1$.}
\label{fig:7}
\end{figure}

The spectrum of the microwave signal that is used to probe the resonator is shown in Fig.\,\ref{fig:7}(a), which is also helpful to frame the quantities described in the text and in Table \ref{tab:1}. Fig.\,\ref{fig:7}(b) shows the resonator response measured with a VNA.
The left sideband is on-resonance with $\omega_r$, and thus interacts with the resonator, while the right sideband remains unaffected. The signal at the carrier frequency $\omega_0$ is suppressed by roughly five orders of magnitude.
This is obtained with a carrier power $\simeq 15\,\mathrm{dBm}$ and frequency $\omega_0/2\pi \simeq 6.16\,\mathrm{GHz}$, while the modulation amplitudes $a_1, a_2$ are of order $50\,\mathrm{mV}$ and frequencies are $\omega_s/2\pi \simeq 10\,\mathrm{MHz}$. The IQ mixer is balanced with DC offsets in order to minimise the amplitude of the carrier at the microwave output by more than four orders of magnitude.
This signal is gradually attenuated by $-91\,\mathrm{dB}$ and sent to the resonator. 

\section{Experimental setup for demonstration of phase and amplitude noise rejection}

\begin{figure}[b!]
\includegraphics[width=.4\textwidth]{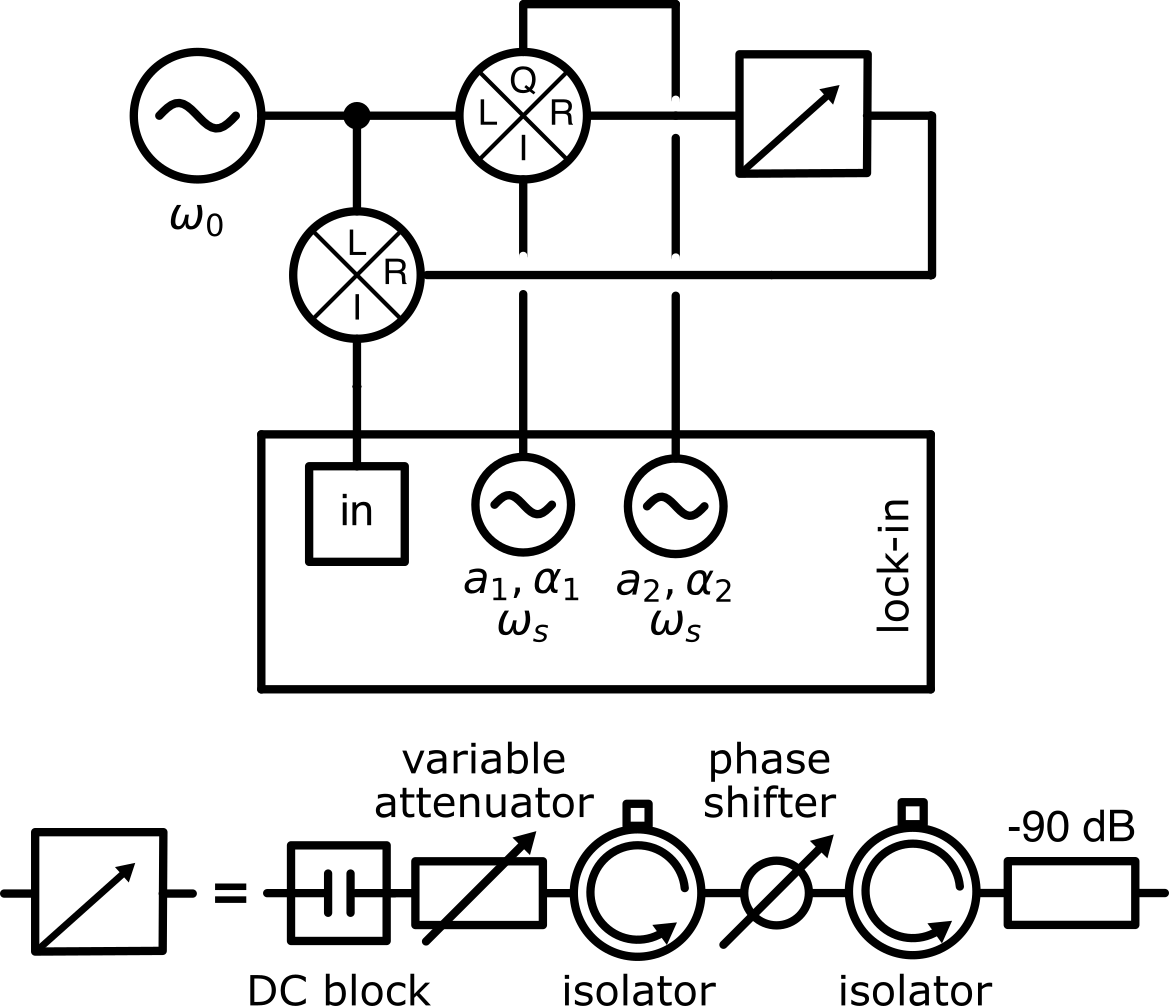}
\caption{Schematic of the experimental setup to demonstrate common phase and amplitude rejection.}
\label{fig:8}
\end{figure}

The experimental setup for the demonstration of phase and amplitude noise rejection found in Sec.\,\ref{sec:noiserej} is displayed in Fig.\,\ref{fig:8}. A voltage-controlled variable attenuator and a variable phase shifter were placed in series after the microwave output of the IQ mixer. A DC block was used to prevent any DC leakage from the IQ mixer that would affect the value of either attenuation or phase arising from these components. An isolator was placed after each component to prevent the formation of standing waves.

For the common-mode phase noise measurements, attenuation arising from the phase shifter was calibrated using a single sideband configuration. A small voltage-dependent attenuation of the phase shifter was compensated with the variable attenuator. A common-mode amplitude variation was achieved by varying both amplitudes $a_1$ and $a_2$ of the signals coming from the lock-in amplifier. 

The different working points in the IQ plane as displayed in Fig.\,\ref{fig:4} were obtained for different amplitudes  $a_2$ ranging from 0\,mV to 80\,mV and phases $\alpha_2$ from 0 to 360\degree\ of the Q-input of the IQ mixer for fixed amplitude  $a_1$\,=\,10\,mV and fixed phase $\alpha_1$. 

\section{Power dependence of resonator noise}
\label{app:power}
\begin{figure*}[t!]
\includegraphics[width=.9\textwidth]{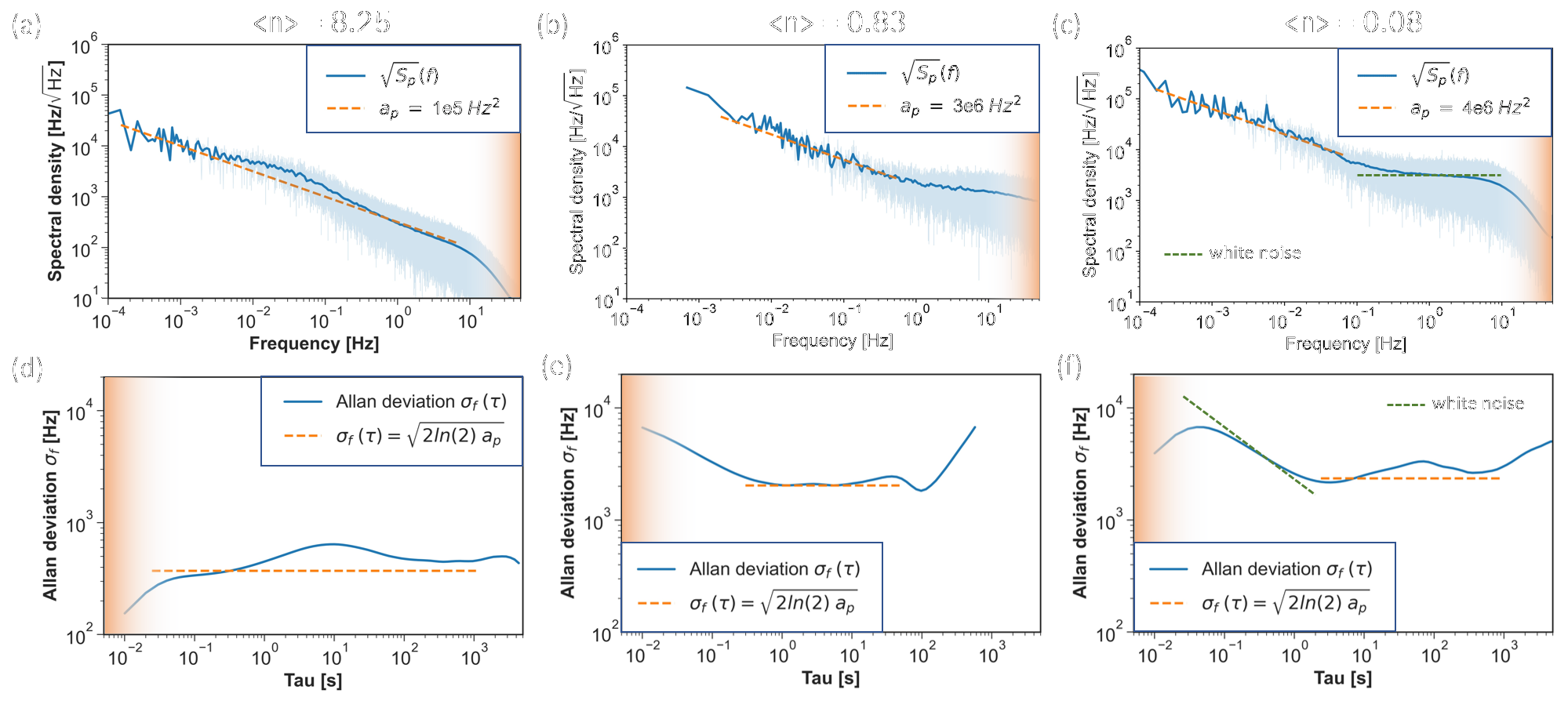}
\caption{Measured spectral densities (a, b, c) and Allan deviations (d, e, f) of the system noise for different probing power as indicated by the average photon numner $\langle n \rangle$ above the plots. }
\label{fig:9}
\end{figure*}

We studied the frequency noise of the resonator for different probing powers and otherwise same parameters as in the main text. The resulting noise spectral densities (SD) and Allan deviations are shown in Fig.\,\ref{fig:9}.
The measurements reported in the main text are (a) and (d), both showing excellent agreement with the expected 1/f behavior up to a timescale of several hours. The reduced probing power, used in (b), (e), (c) and (f), decreases the signal-to-noise ratio of the interferometer, therefore measurement noise (white noise) starts to dominate over the one from the resonator.
For both measurements at weaker probing power the Allan deviation shows a slope closer to -0.5 (green dashed line) at small $\tau$ values which is expected for white noise.

\section{Guide of operation}
\label{app:guide}
Here, we detail a practical guide to setup an interferometric readout like the one described in this work. In this context, Fig.\,\ref{fig:2} is taken as a reference.
\begin{enumerate}
    \item Take a microwave source and tune its frequency (carrier) $\omega_0$ close to the one of the DUT.
    \item The carrier is then split, and the two outputs are connected to the LO ports of an IQ mixer and a down-conversion mixer. In the former, the in-phase and quadrature inputs are wired to two low-frequency oscillators which create sidebands and have the same frequency $\omega_1=\omega_2$ (but tunable phase $\alpha_1$ and $\alpha_2$ and amplitude $a_1$ and $a_2$).
In our setup these are the reference outputs of the lock-in amplifier and can therefore also be used for phase-sensitive detection of the down converted output signal of the SMI. The amplitude of the probe signal can be varied by changing $a_1$ and $a_2$, or by attenuating the entire input signal.
    \item At this point, using similar initial values of the two amplitudes and phases, it is convenient to connect the resulting signal to a spectrum analyser and check the sidebands. The DC offsets of the low-frequency oscillators can eliminate the residual carrier at $\omega_0$ by balancing the mixing, and yield a spectrum like the one of Fig.\,\ref{fig:7}(a).
This is the input signal that is sent to the DUT, so one should take care that only one sideband is on-resonance with the spectral features of the DUT. The output signal of the SMI is down-converted in the second mixer, and its phase and amplitude detected with the lock-in amplifier.
    \item With this the setup is fully wired, and measurements can commence.
The first thing to test is that one sideband is actually on resonance with the DUT, which can be done for example by sweeping $\omega_0$ and resolving the spectral features of the DUT on the lock-in output. This spectroscopic measurement can also be fit to the resonator lineshape (measured e.\,g. with a VNA) to calibrate the system. Once the resonance of the DUT is found, the amplitude-noise rejection configuration is simply obtained by tweaking $a_2$ and $\alpha_2$ to get a zero in the amplitude $\sqrt{x^2+y^2}$ of the output signal, as is shown in Fig.\,\ref{fig:4}(a).
Before starting measurements, the input signal is checked again and the mixer re-balanced such as to account for intermittent adjustments of the oscillator's parameters.
\end{enumerate}
The interferometer is ready to measure in the amplitude-noise rejection configuration. The simplest way to switch to phase-noise rejection requires a phase-shifter between the splitter and the LO (carrier or local oscillator) input of the down-conversion mixer. The interferometer can be configured to the amplitude-noise rejection following the previous procedure, and by applying a phase shift of $\pi$ one ends up in the phase-noise rejection configuration.

\bibliographystyle{apsrev4-2}
\bibliography{sidebandInterferometer}

\end{document}